\documentclass{ws-jai}
\usepackage[flushleft]{threeparttable}
\begin{document}

\catchline{}{}{}{}{} % Publisher's Area please ignore

\markboth{Samuel Gordon}{An Open Source, FPGA-based LeKID readout for BLAST-TNG: Pre-flight Results}

\title{An Open Source, FPGA-based LeKID Readout for BLAST-TNG: Pre-flight Results}

\author{Samuel Gordon$^{a}$, Brad Dober$^{b}$, Adrian Sinclair$^{a}$, Samuel Rowe$^{c}$, Sean Bryan$^{a}$, Philip Mauskopf$^{a}$, Jason Austermann$^{d}$, Mark Devlin$^{b}$, Simon Dicker$^{b}$, Jiansong Gao$^{d}$, Gene C. Hilton$^{d}$, Johannes Hubmayr$^{d}$, Glenn Jones$^{e}$, Jeffrey Klein$^{b}$, Nathan P. Lourie$^{b}$, Christopher McKenney$^{d}$, Federico Nati$^{b}$, Juan D. Soler$^{f}$, Matthew Strader$^{g}$, Michael Vissers$^{d}$}

\address{
$^{a}$School of Earth \& Space Exploration, Arizona State University, 781 E Terrace Mall, Tempe, AZ 85287, USA \\
$^{b}$Depart of Physics \& Astronomy, University of Pennsylvania, 209 S 33rd St, Philadelphia, PA 19104, USA \\
$^{c}$School of Physics \& Astronomy, Cardiff University, Queen's Buildings, The Parade, Cardiff, CF24 3AA, Wales, UK\\
$^{d}$National Institute of Standards \& Technology, 325 Broadway, Boulder, CO 80305 \\
$^{e}$Department of Physics, Columbia University, 538 W 120th St, New York, NY 10027, USA \\
$^{f}$Department of Physics \& Astronomy, University of British Columbia, 6224 Agricultural Road, Vancouver, BC, Canada\\
$^{g}$Department of Physics, University of California, Santa Barbara, CA 93106, USA\\
}

\maketitle

\begin{history}
\received{(to be inserted by publisher)};
\revised{(to be inserted by publisher)};
\accepted{(to be inserted by publisher)};
\end{history}

\begin{abstract}
We present a highly frequency multiplexed readout for large-format superconducting detector arrays intended for use in the next generation of balloon-borne and space-based sub-millimeter and far-infrared missions. We will demonstrate this technology on the upcoming NASA Next Generation Balloon-borne Large Aperture Sub-millimeter Telescope (BLAST-TNG) to measure the polarized emission of Galactic dust at wavelengths of 250, 350 and 500 microns. The BLAST-TNG receiver incorporates the first arrays of Lumped Element Kinetic Inductance Detectors (LeKID) along with the first microwave multiplexing readout electronics to fly in a space-like environment and will significantly advance the TRL for these technologies. After the flight of BLAST-TNG, we will continue to improve the performance of the detectors and readout electronics for the next generation of balloon-borne instruments and for use in a future FIR Surveyor.
\end{abstract}


\section{Introduction}

Millimeter and sub-millimeter wave astronomy provides a view of the early universe through the Cosmic Microwave Background (CMB), high-redshift galaxies and star formation with the Cosmic Infrared Background (CIB), and magnetic fields and local star formation through Galactic polarimetry. Several current and planned experiments aim to map the CMB polarization anisotropies. These include ground-based ({\it e.g.} ACT-Pol \cite{niemack10}, SPT-Pol \cite{austermann12} and POLARBEAR \cite{kermish12}), balloon-borne ({\it e.g.} SPIDER \cite{crill08}, EBEX \cite{reichborn-kjennerud10} and PIPER \cite{chuss10}), as well as space-based experiments ({\it e.g.} PLANCK \cite{ade13} and LiteBIRD \cite{matsumura14}). The trend for each of these experiments is a push towards increased sensitivity and scanning speed, which entails higher pixel count arrays. The larger arrays in turn require readout electronics with increased multiplexing factors to reduce the power consumption per pixel. We have recently developed and demonstrated an FPGA-based, room temperature, highly multiplexed readout platform for multi-kilopixel arrays of superconducting detectors which will fly on the Next Generation Balloon-Borne Large Aperture Sub-millimeter Telescope (BLAST-TNG). 

BLAST-TNG is a polarimeter that will measure emission from cosmic dust in nearby giant molecular clouds (GMCs) over three narrow frequency bands centered at 250, 350 and 500 $\mu$m \cite{dober14,galitzki16}. Its targets include well known cloud complexes including Ophiuchus, Lupus and Vela, along with several infrared dark clouds. The flight is scheduled to take place over 30 days at McMurdo Station, Antarctica, in December, 2017. The BLAST-TNG receiver is based on Lumped Element Kinetic Inductance Detectors (LeKID) \cite{doyle08}. LeKID detectors are superconducting resonators that change their resonant frequency in response to the absorption of incident photons that have enough energy to break apart superconducting Cooper pairs. Arrays of  LeKIDs are highly multiplexable and represent a promising detector technology for the next generation of balloon-borne and space-based Far-Infrared (FIR) missions. 

The readout, which is the subject of this paper, is based around a set of five, 1024-channel digital filterbank spectrometers. For this readout system architecture, the number of channels in the filterbank corresponds directly to the maximum number of resonators that can be read out simultaneously. It builds on previous KID demonstrator instruments that have used Field Programmable Gate Array (FGPA) platforms, which enable the high speed digital signal processing (DSP) techniques required for multiplexed readout ({\it e.g.} MUSIC \cite{golwala13}, ARCONS \cite{mchugh12}, NIKA \cite{monfardini14}, MUSTANG-2 \cite{dicker14}, MAKO \cite{swenson12}) and SPACEKIDS \cite{vanrantwijk16}. It is based on the second generation Reconfigurable Open Architecture Computing Hardware (ROACH-2), an open source DSP platform developed by the Collaboration for Astronomical Signal Processing and Electronics Research \cite{Werthimer11, hickish2016}. BLAST-TNG will mark the first time that such a readout platform has been flown on a long duration balloon flight. 

The system was recently demonstrated at the National Institute of Standards and Technology (NIST), where it simultaneously read out several hundred LeKID detectors from the BLAST-TNG 250 $\mu$m prototype array. During this test it was determined that the read-noise is below the detector noise threshold. In the following sections, we provide an overview of the readout electronics, firmware and software development, payload integration and preliminary results of system noise characterization which will continue during the months leading up to the planned December, 2017 launch. 

\section{System Requirements}

The basic requirements of the readout system are that it must be able to read out the BLAST-TNG LeKID detectors at a rate determined by the telescope scan speed, while having a noise contribution less than that of the detectors and other receiver systems. The readout noise has two dominant components, which are white noise and flicker noise (1/f). The intrinsic white noise floor of the BLAST-TNG detectors sets a white noise readout requirement of $\lesssim{-100}$~dBc/Hz in the phase direction. Since the detectors will be read out using frequency modulation, the requirement on the noise in the amplitude direction is not as stringent. 

The detector 1/f knee, which is $\sim$0.5~Hz for all three science bands, sets a lower limit on the telescope scan speed and readout frequency that will permit for adequate mapping of BLAST's science targets. For BLAST-TNG, the desired readout frequency is between 244 - 488~Hz, where each channel outputs a 64-bit complex valued sample at this rate. Regardless of the number of active channels, the data packet size is fixed at 8192 bytes, resulting in a data rate of $\sim$4~MB/s. The data must be continuously stored to disk during flight, and time stamped for synchronizing with telescope pointing information. 

\begin{wstable}[h]
\caption{Detector Counts} \label{aba:tbl1}
\begin{tabular}{l|ccc} \toprule
	   Array & 250~$\mu$m & 350~$\mu$m & 500~$\mu$m \\ \colrule
	   Readout Modules & 3 & 1 & 1 \\ 
	   Number of Tones per Module & 612 & 950 & 544 \\ \hline 
\end{tabular}
\end{wstable} 

The readout must meet its noise requirements at the highest multiplexing factor set by the pixel count of each detector array. BLAST-TNG's $\sim$3330 detectors are divided between five independent readout modules: One each for the 350 and 500~$\mu$m arrays, and three for the 250~$\mu$m array. Each module contains a ROACH-2 board, DAC/ADC board, and a set of RF front-end electronics (see Section \ref{sec:electronics}). Although the 250~$\mu$m array could have been distributed amongst two modules, the array was designed as a set of three identical rhombuses in order to ease their fabrication. The 950 pixels of the 350~$\mu$m array sets the most stringent multiplexing requirement on a single module. 

By virtue of launching in the Antarctic summer, the primary requirement regarding power dissipation is that the system must be able to passively cool in the extreme environmental conditions present at altitude ($\sim$35~km). Each module has a power budget of $\lesssim{100}$~W (see Section \ref{sssec:thermal}). The primary system requirements are listed in Table \ref{aba:tbl2}. 

\begin{wstable}[h]
\caption{Readout Requirements} \label{aba:tbl2}
\begin{tabular}{@{}cccc@{}} \toprule
RF bandwidth & White Noise & Flicker Noise & Readout frequency \\ \colrule
512~MHz\hphantom{0} & \hphantom{0}$\lesssim{-100}$~dBc/Hz &\hphantom{0}$\lesssim{0.5}$~Hz & 244-488~Hz \\ \hline \\
Power dissipation\hphantom{0} & \hphantom{0}Timing accuracy & \hphantom{0}Tone resolution& Num channels per module \\ \hline \\
$\lesssim{100}$ W\hphantom{0} & $\lesssim{2}$~ms & $\leq{1000}$~Hz\hphantom{0} & $\gtrsim{500}$\hphantom{0} \\ \botrule
\end{tabular}
\label{aba:tbl2}
\end{wstable}

\section{Hardware and Electronics}

The following two sections provide an overview of the readout hardware. We will cover the electronics, and the enclosure which houses them while managing their power supply, temperature and signal routing during flight.

\subsection{Electronics}
\label{sec:electronics}

\begin{figure}[h]
\centering
\includegraphics[width=0.9\textwidth]{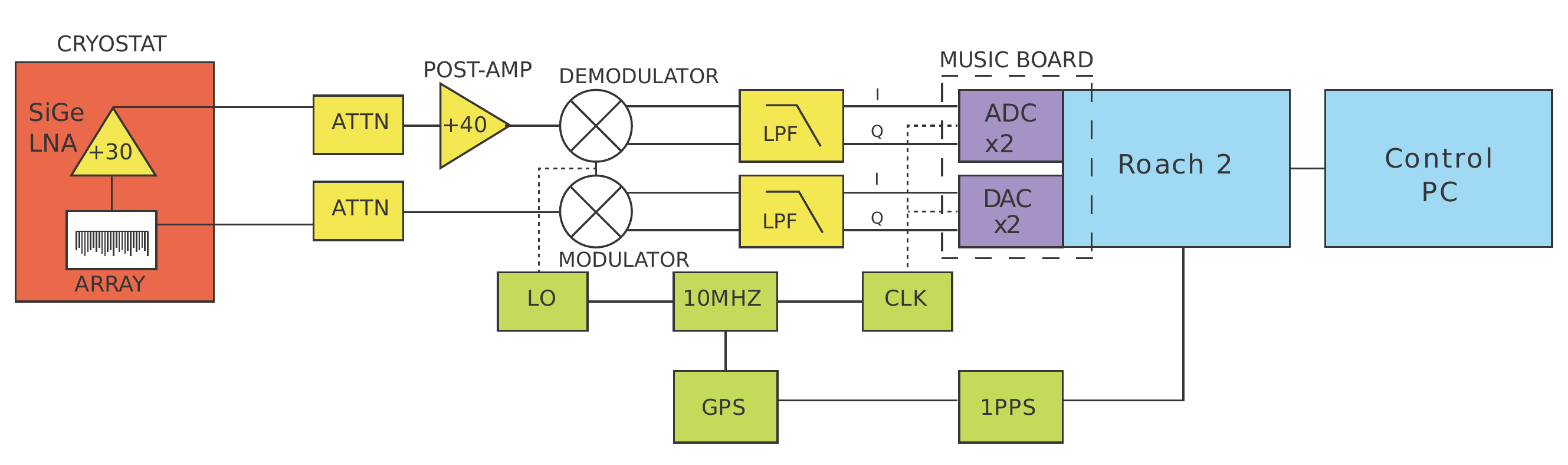}
\caption[Block Diagram of the BLAST-TNG Readout Electronics.]{A schematic of the BLAST-TNG readout electronics.\label{aba:fig1}}
\end{figure}

The readout electronics, shown conceptually in the top panel of Figure \ref{aba:fig1} for a single readout module, consists of a ROACH-2 Virtex-6 FPGA board coupled to a MUSIC DAC/ADC board \cite{duan10}. An embedded processor (AMCC PowerPC 440EPx, hereafter PPC) acts as an interface between the FPGA and data acquisition (DAQ) or flight computer (FC). 

The PPC runs a daemonized Karoo Array Telescope Protocol (KATCP) server, which facilitates communications between the DAQ and FPGA\footnote{KATCP has been developed by the Square Kilometer Array South Africa (SKA SA) collaboration for use on their CASPER hardware-based correlators and beam formers. See
\url{https://casper.berkeley.edu/wiki/KATCP}}. The MUSIC board includes two 12-bit 550 MSample/s ADC chips\footnote{ADS54RF63, Texas Instruments Inc.}, and two 16-bit 1000~MSample/s DAC chips\footnote{DAC5681, Texas Instruments Inc.}. Data can be streamed to the DAQ via two 1GbE Ethernet tranceivers, one each for the PPC and FPGA. The PPC tranceiver is used for commanding and reading back small quantities of diagnostic data from the FPGA, while the FPGA tranceiver streams User Datagram Protocol (UDP) data packets to the DAQ. A dual channel Valon\footnote{5008 Dual-Frequency Synthesizer, Valon Technology Inc. 750 Hillcrest Drive Redwood City, CA 94062} analog synthesizer provides the 512~MHz sampling clock (CLK) for the MUSIC board, as well as the tunable local oscillator (LO) for an external quadrature modulator\footnote{AM0350A, Polyphase Microwave. 1983 Liberty Dr, Bloomington, IN 47403} and demodulator\footnote{AD0105B, Polyphase Microwave.}. The FPGA and DSP runs at 256~MHz, which it derives from the sampling clock.

The analog front-end performs up and down-conversion of a baseband frequency comb spanning -256 - 256~MHz output by the ROACH-2 and MUSIC board. The radio frequency range (RF) containing the resonant frequencies of the LeKID detectors is 500 - 1012~MHz. Digitally programmable\footnote{RUDAT-6000-30, Mini-Circuits. 13 Neptune Ave Brooklyn, NY 11235} attenuators are used at the output of the modulator and at the input of the demodulator to match the total frequency comb power to the optimal detector tone power and full-scale dynamic range of the ADCs. The RF input attenuator is preceeded by a room temperature amplifier\footnote{ZKL-1R5+, Mini-Circuits}, which provides +40 dB of gain. The analog synthesizer and attenuators are controllable from either of two flight computers (FCs) via a BeagleBone Green single board computer\footnote{Seed Development Limited. 1933 Davis Street, Suite 266, San Leandro, CA 94579}. 

\subsection{ROACH-2 Motel}\label{sec:roachmotel}

The ROACH-2 Motel is a custom enclosure that houses the set of five ROACH-2 modules and distributes power to each of the electronic components listed in \ref{sec:electronics}. To allow for continuous operation at high altitude, the enclosure provides a thermal link to the inner frame of the balloon gondola, through which the power generated by the electronics can dissipate and then be radiated to space.

\begin{figure}[h]
\centering
\includegraphics[width=0.5\linewidth]{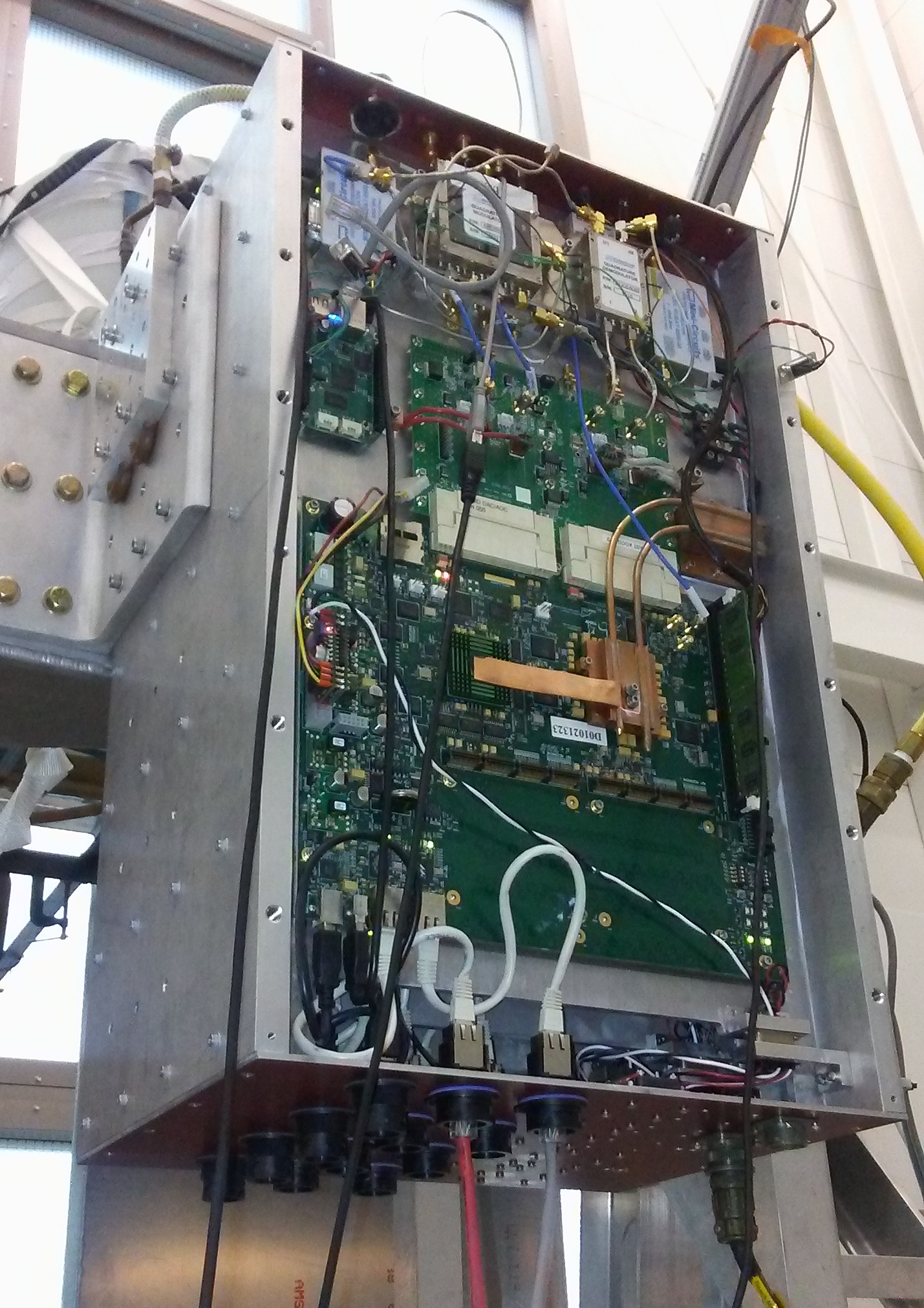}
\caption[ROACH-2 Motel Components]{The ROACH-2 flight enclosure, containing four out of five modules, mounted to the inner frame of the BLAST-TNG gondola. The overall dimensions of the ROACH-2 motel with all five modules is 14.75" x 2.75" x 24". The FPGA is heat sunk via a custom heat pipe assembly. The PPC is heat strapped to the FPGA heat sink, while the ADCs and DACs are strapped directly to the aluminum backing plate. Also shown are the input and output attenuators, second stage amplifier, BeagleBone, quadrature modulator and demodulator. The Valon synthesizer is mounted underneath the modulator. \label{aba:fig2}}
\end{figure}

Table~\ref{aba:tbl3} lists the power dissipation of major components on the ROACH-2 and MUSIC board as measured during testing in RF `loopback' mode. In RF loopback mode, the ROACH-2 is programmed with a probe comb of 1000 tones, which are looped back continuously from the DACs to the ADCs, through the front end electronics.
\begin{wstable}[h]
\caption[ROACH-2 Power Dissipation]{ROACH-2 power dissipation shown by largest contributors and displayed in watts. The FPGA dissipates nearly as much as the rest of the ROACH-2 board combined. \label{aba:tbl3}}
\begin{tabular}{ ccc } \toprule
	   Component & Quantity & Power Dissipation (W) \\ \colrule
	   FPGA & 1 & 30 \\ 
	   Power PC & 1 & 5.0 \\ 
	   RAM & 1 & 4.4 \\
	   ADCs & 2 & 2.6 \\
       QDR & 4 & 1.8 \\
       PHY & 2& 1.0 \\ 
       DACs & 2 & 0.5 \\	\botrule 
\end{tabular}
\end{wstable}

The ROACH-2 circuit components are all of the industrial variety and are rated for operation at over 85$^\circ$C. The ROACH-2 Motel needs to properly heat sink the FPGA, PPC and ADCs to prevent them from exceeding this temperature and failing during flight. The large amount of power dissipated by the FPGA necessitates a low thermal resistance path to keep the temperature within an acceptable range. Two 5~mm diameter water filled, sintered copper wick heat pipes\footnote{HP-HD05DI25000BA, Enertron Inc. 90 N William Dillard Dr., Suite 121 Gilbert, AZ 85233} are installed into custom heat sinks via bismuth tin (BiSn) solder paste. BiSn is used because its relatively low melting point of 138$^{\circ}$C allows it to flow before damaging the heat pipe. 

A thermal conductivity calculation program\footnote{I. Advanced Cooling Technologies. Heat pipe calculator, 2016.} was used to analyze the heat pipe. The results confirm that two heat pipes in their existing configuration are rated to carry $\sim$30~W safely at $\sim$40$^{\circ}$C. The PPC is heat-sunk to the FPGA heat sink via a conventional 1/64" thick, 2/3" wide copper strap. The ADCs are heat strapped directly to the 1/4" Aluminum backing plate by two 10 AWG copper wires. The DACs, which run cooler than the ADCs, are strapped to the backing plate by a single 14 AWG copper wire. The ROACH-2, MUSIC board and analog electronics for each readout module are mounted to separate aluminum backing plates (Figure \ref{aba:fig2}). The Beaglebone dissipates an inconsequential amount of power, and is placed on aluminum standoffs.

The backing plate for each readout module is mounted between two 5/8" thick aluminum side panels via via fourteen 8-32 threaded screws that provide compression contacts. Heat from the backing plate is conducted through each of the side panels and into two 8" x 5" x 1/4" right angle brackets which mount the entire Motel to the inner frame of the gondola.

\subsubsection{Thermal Verification}\label{sssec:thermal}

The ROACH-2 Motel thermal design was vetted by running multiple stress tests inside a vacuum chamber. Since the inner frame is unable to fit inside the vacuum chamber, two water heat exchangers are mounted on top of the brackets, and kept fixed at the expected temperature of the inner frame during flight. During the vacuum test, the four completed systems are powered up, and the firmware is uploaded to the FPGA. The readout software is run continuously in an RF electronics loopback mode, where the RF output is connected directly to the RF input. The component temperatures after the Motel reached equilibrium are displayed in Table \ref{aba:tbl4}. All components behaved as expected, and the hottest elements, the PPCs, were still $29^{\circ}C$ below their maximum allowable temperature.

\begin{wstable}[h]
\caption[ROACH-2 Motel Component Temperatures Measured Under Vacuum]{Various temperatures of ROACH-2 Motel components measured under vacuum. Every component is well within thermal tolerances, and the highest temperature component, the PPC, is $29^{\circ}$C below its maximum allowable temperature.}

\label{aba:tbl4}
\begin{tabular}{c | cccc} \toprule
Component & PPC & FPGA & Evaporator & Condenser \\ \colrule
Temperature ($^{\circ}$C) & 56 & 41 & 37 & 33 \\
\hline \\
Component & Inlet & Outlet & ADC & DAC \\ \colrule
Temperature ($^{\circ}$C) & 44 & 36 & 39 & 35 \\ \hline
\end{tabular}
\end{wstable}

The results of the thermal vacuum test were used to calibrate the ROACH-2 Motel thermal model. These data were incorporated into a full simulation of the BLAST-TNG gondola thermal environment during flight. This model is designed and simulated with Thermal Desktop\footnote{C\&R Technologies. 2501 Briarwood Dr, Boulder, CO 80305}\textsuperscript{\textregistered}. Thermal Desktop\textsuperscript{\textregistered} creates a node and conduction network from a CAD model, interfaces with SINDA/FLUINT \cite{Cullimore98} which solves the heat transfer equations, and interprets and displays the results.

\begin{figure}
\centering
\includegraphics[width=1.0\linewidth]{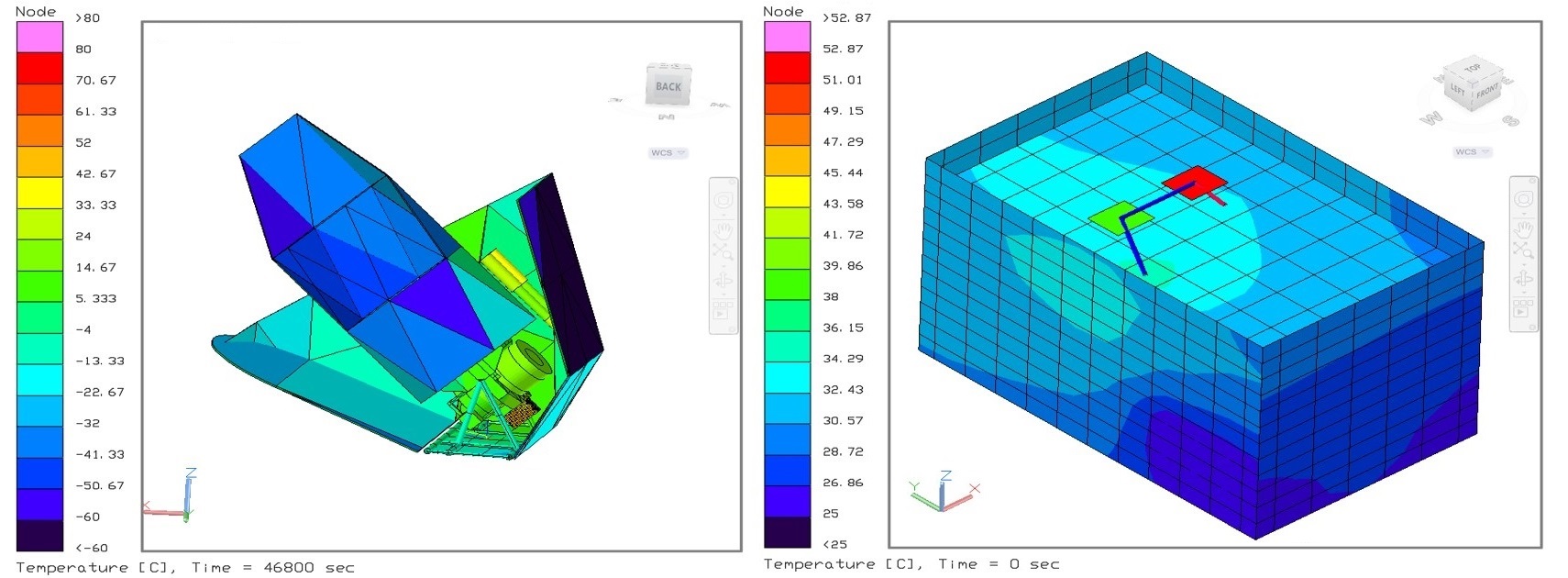}
\caption[Thermal Simulations of the Gondola at Float]{Left: The Thermal Desktop model that was used to cross-check the measured temperatures in the vacuum chamber test, which are listed in Table \ref{aba:tbl3}. Right: The thermal simulation of the gondola at float, with the ROACH-2 Motel attached. Simulations suggest the ROACH-2 Motel will passively cool to $\sim$$40^{\circ}$C. \label{aba:fig3}}
\end{figure}

The Systems Improved Numerical Differencing Analyzer and Fluid Integrator (SINDA/FLUINT) is the NASA standard for computationally simulating heat transfer and fluid flow networks. The results of a simulated flight, shown in Figure \ref{aba:fig3}, which incorporates all critical hardware components and the proper sun-shield design, suggest that the inner frame of the gondola is able to safely conduct and radiate away all of power generated by the ROACH-2 Motel, provided its sides are painted white to reflect incident solar radiation, while efficiently radiating in the infrared.

\subsubsection{Power and Interfacing} 

In addition to providing a mounting point and thermal link for all of the electronics, the ROACH-2 Motel also distributes all of the required power and signals. Each module front panel has SMA ports for a 10~MHz reference, a pulse-per-second (PPS) sync, an (unused) external LO, RF input and output, and a spare. There is both a power and reset switch which must be `armed' using a third switch to ensure that modules aren't accidentally powered down during operation. Finally, there is an Ethernet port for connecting to each BeagleBone. 

The back panel of each module features a four pin military connector which receives the 28 VDC supply from the gondola's power distribution system. This power is split out to several Vicor DC-DC converters which are mounted on the Motel's backing plate and provide the $\pm$5 and 12 VDC supplies for the I/Q modulator, demodulator, attenuators, Valon synthesizer and second-stage amplifier. The 28 VDC power is fed into a PicoPSU ATX power supply for the ROACH-2 board, and +12 and $\pm$5 VDC power for each module is fed from the Vicor DC-DC converters via a 12-pin connector. Each back panel also contains Ethernet ports for the PPC and FPGA, as well as a USB port for interfacing with the PPC to be used for diagnostics and debugging.

\section{BLAST-TNG ROACH-2 Firmware Overview}

The BLAST-TNG ROACH-2 firmware running on each of the five modules is a 1024-channel filter bank spectrometer covering 512~MHz of instantaneous baseband bandwidth. This number of channels is set by BLAST-TNG's system requirements, and does not constitute an upper limit on the FPGA resources. The large frequency domain multiplexing factor is accomplished by performing simultaneous digital down-conversion (DDC) of each channel using a look-up-table (LUT) of digital LO waveforms, a technique which has been successfully implemented by the ARCONS experiment \cite{mchugh12}.

\begin{figure}[h]
\centering
\includegraphics[width=0.9\textwidth]{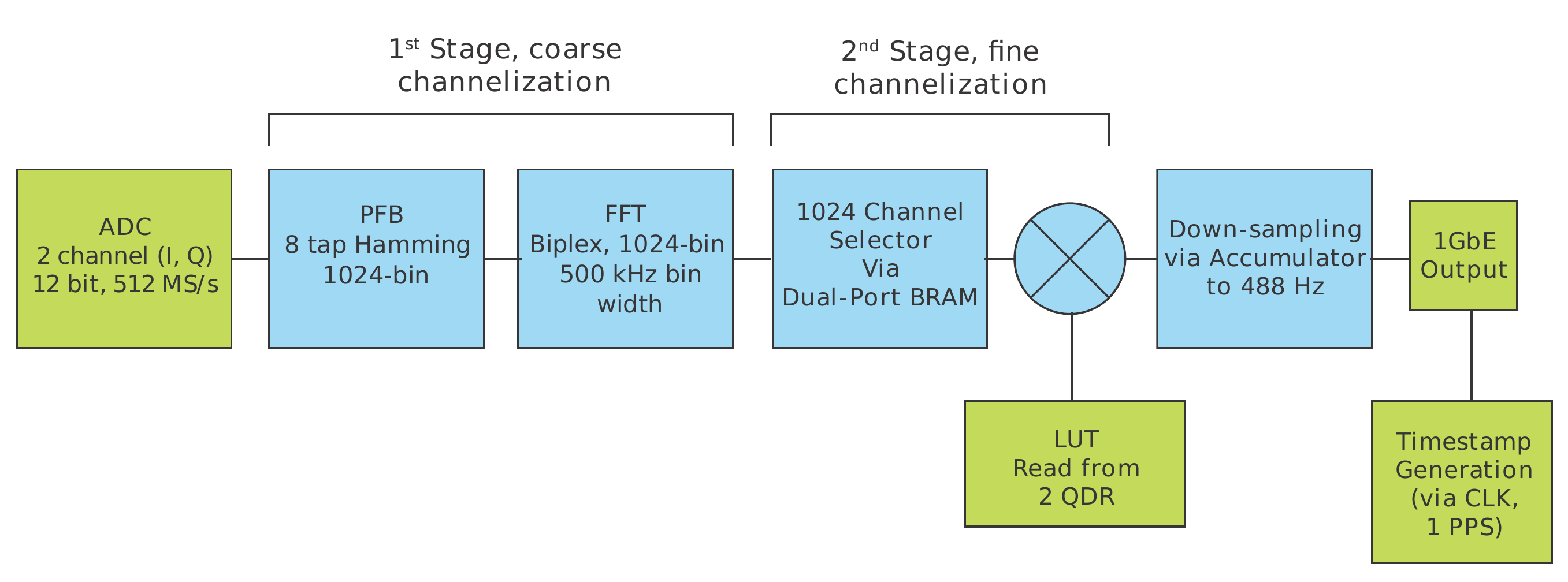}
\caption[Schematic of the firmware DSP chain.]{A schematic of the DSP chain.\label{aba:fig4}}
\end{figure}

The firmware is written using the MATLAB\footnote{Mathworks, 1 Apple Hill Drive Natick, MA 01760} / Simulink\footnote{Mathworks} / System Generator\footnote{Xilinx ISE 14.7 Design Suite} / EDK\footnote{Xilinx Embedded Development Kit} (MSSGE) Toolflow developed by the CASPER collaboration. CASPER 'snap' blocks allow for pre-specified amounts of data from the data stream to be saved in block RAM on the FPGA at key points in the DSP chain. This data can then be converted into figures of merit to be used for making on-the-fly adjustments to either the RF electronics or baseband frequency combs. Critical parameters, including the readout frequency, may be adjusted during operation using software inputs.

The signal processing chain is composed of two parts: probe tone synthesis and analysis, which are described in detail below. A block diagram of the signal processing chain is shown in the bottom panel of Figure \ref{aba:fig4}. Its purpose is to generate a comb of microwave carrier tones (probe tones) which are continuously fed to each of the LeKID detectors on a single microwave feed-line. After interacting with the resonators,
the probe tone comb is digitized by the ADCs. The time ordered data then is then filtered, Fourier-transformed, digitally down-converted, downsampled to 488 Hz, packetized and logged to disk. Although each DAC and ADC has a Nyquist-limited bandwidth of 256~MHz, the use of separate chips to process the in-phase and quadrature components of a probe comb containing both positive and negative frequencies results in an effective system bandwidth of 512~MHz. For a detailed discussion of I/Q sampling with separate chips, see \citet{Jones10}. In the following, we divide each firmware operation into two distinct stages: Probe comb synthesis and analysis. 

\subsection{Probe Comb Synthesis}
\subsubsection{Carrier Waveform Buffer}

The baseband probe comb LUT buffer occupies two of the ROACH-2's four quad-data-rate (QDR) SRAM chips\footnote{Cypress, CY7C2565KV18}, hereafter designated QDR$_{I}$ and QDR$ _{Q}$. This buffer includes both the LUT to be processed by the DAC (DAC LUT), as well as the LUT to be used for DDC (DDC LUT). The quadrature waveform buffers are generated in software (see Section \ref{sec:software}) prior to being uploaded to the QDRs, and contain frequencies between -256 and +256 MHz. Each LUT comprises 2$^{21}$ signed 16-bit time ordered samples. Dividing this length into the 512 MHz DAC sampling frequency yields a frequency resolution of 244.14 Hz for the $I$ and $Q$ time series, respectively. Combining the two quadratures results in a probe tone frequency resolution of 122.07 Hz. 

The QDR is logically accessed as 2$^{19}$ addresses x 16 bytes. The KATCP protocol is used to upload the LUT data to QDR RAM. To facilitate uploading the two LUTs to each QDR, the $I$ and $Q$ components are interwoven into two separate LUTs (LUT$_{I}$, LUT$_{Q}$) of 2$^{22}$ time ordered samples each. For example, the order of samples contained within LUT$_{I}$ is:

\begin{equation} I_\textup{DAC}^{1}, I_\textup{DAC}^{0}, I_\textup{DDC}^{1},I_\textup{DDC}^{0}, ... , I_\textup{DAC}^{n}, I_\textup{DAC}^{n -1 }, I_\textup{DDC}^{n}, I_\textup{DDC}^{n - 1} \end{equation} 
where superscripts refer to even and odd numbered samples, and each value is 16-bits wide.

After the LUTs have been uploaded to the QDRs, they are read from the QDR buffers, sliced into their original 16-bit components and recast as a fixed-point 16.15 number\footnote{In this fixed-point notation, the first number represents the total bit width, with the second number being the radix point. All numbers in this notation are assumed to be signed unless otherwise noted.}. On each clock cycle, four consecutive samples are read out from QDR$_{I}$ (QDR$_{Q}$): $I_\textup{DAC}^{1}, I_\textup{DAC}^{0}, I_\textup{DDC}^{1},I_\textup{DDC}^{0}$ (same for $Q$). The DDC LUT samples are sent directly to the DDC section of the firmware, while The DAC LUT samples are presented to the two DACs, which generate the I and Q components of the baseband signal that drives the quadrature modulator. Since the DACs are clocked at twice the rate of the FPGAs, two consecutive samples ({\it e.g.} $I_\textup{DAC}^{1}, I_\textup{DAC}^{0})$ are processed on each FPGA clock cycle. To ensure proper synchronization of each quadrature component, the DAC is synchronized by a pulse which also resets the QDR address counter to zero. The analog waveform is then upconverted to RF, and passed through the detector feedline for modulation inside the cryostat.    

\subsection{Analysis}
\subsubsection{Digitization}

The downconverted and re-digitized baseband frequency comb contains frequency information spanning -256 to 256~MHz. The two quadrature ADCs are synchronized in firmware by the same software register as the DACs. Each ADC outputs two consecutive time samples per FPGA clock cycle: I$_\textup{ADC}^{1}$, I$_\textup{ADC}^{0}$ (Q$_\textup{ADC}^{1}$, Q$_\textup{ADC}^{0}$). The time ordered $I/Q$ pairs are concatenated and sent to the first stage of channelization. Snap blocks are utilized at this stage to store some of the ADC time stream, which is downloaded to the DAQ. The data is used to calculate the RMS voltage measured at the ADC for comparison to the maximum input range of 2.2~V (p-p), corresponding to an input power of 10.32 dBm. The input attenuator is then adjusted in steps of 0.5 dB to bring the measured ADC input power within range of the full scale, while avoiding saturation.

\subsubsection{Coarse Channelization}

The digitized signal is channelized using a polyphase filterbank (PFB). The PFB is implemented using the CASPER pfb\_fir block and the biplex\_fft block. The pfb\_fir
implements an 8-tap Hamming window, which reduces spectral leakage and scalloping\footnote{For details on the CASPER PFB, see: \url{https://casper.berkeley.edu/wiki/The_Polyphase_Filter_Bank_Technique}}. On each clock cycle, the biplex FFT receives two consecutive complex time ordered samples, and outputs the complex amplitudes ($i$ and $q$) of two consecutive frequency bins. One 1024-bin FFT is processed every 512 clock cycles, corresponding to an FFT-rate of 500~kHz. A synchronization pulse which is emitted on the last clock cycle before the first valid data of each consecutive FFT is used to synchronize all following stages of the firmware. Since the average individual detector bandwidth is $\sim$50~kHz, several detectors may safely fall within a single FFT bin. Each bin pair output by the FFT is concatenated into a single 72-bit word (4x18 bit) with order $i_{1}$, $q_{1}$, $i_{0}$, $q_{0}$ (lower case $i,q$ are hereafter used to denote FFT-bin time streams), before being stored in block RAM (BRAM) in the FPGA for channel selection. 

\subsubsection{Fine Channelization}

Since some FFT bins will contain multiple carriers, while others remain empty, only the former set of bins requires further channelization. The channel selection logic requires that up to 1024 channels from the FFT bin stream be selected within 512 clock cycles. To manage this while continuously streaming data, a buffered switch is constructed using Xilinx dual-port BRAM blocks. During software synthesis of the $I/Q$ waveform buffers, a list of up to 1024 bins is pre-calculated based on known resonator positions and loaded into a dual-port `bin select' BRAM. The list may consist of any combination of the 1024 available bin indices, including a single bin index repeated 1024 times. If only a subset of bins is required, any unused RAM addresses are initialized to zero. Once the bin list is loaded into RAM, the bins are referred to as channels, with the channel order corresponding to the order of the original list. During operation, the bin indices for two consecutive channels are read out in parallel from the dual-port RAM. Each bin index is halved to represent the clock cycle (`clock address') corresponding to its offset in cycles from the zeroth FFT bin, and these clock addresses are to be used as read addresses for another dual-port RAM containing the bin data. In the data RAM, the contents of two consecutive bins are stored in each address slot. 

In read mode, the contents of two bins are presented at each output port of the dual-port data RAM, the addresses of which are chosen by the clock addresses of the desired bin indices. Out of the four available bins to choose from on each cycle, only one member of each pair is desired. To determine which member of each bin pair to use, the least significant bit of the desired bin index is used to operate a switch that slices the proper bin from each pair. The new pair of bins is then passed through a MUX selector and sent to the first stage of the DDC. To facilitate continuous readout, the bin selector is duplicated into a read branch and a write branch, which together form a buffered switch.

\subsubsection{Digital Down Conversion}

The FFT operates on the digitized ADC time stream once every 512 clock cycles, and therefore any probe tone waveforms that have a period longer than the filterbank FFT length (1024 samples) will exhibit unwanted phase rotation over the course of several FFTs. The result is amplitude modulation (AM) of each FFT bin time stream, where the AM frequency is the beat frequency between the filterbank bin center frequency and the location of a carrier tone within the bin. One approach to circumventing this AM is to use a longer FFT, so that each carrier tone falls very near to a bin center. Previously, the MUSIC \cite{duan10} firmware employed this approach, with a $2^{16}$-point FFT, resulting in bin width of $\sim${7.5}~kHz. Instead, we use digital down conversion to demodulate this residual AM. This concept was previously implemented in the ARCONS firmware \cite{mchugh12}. This technique offers the advantage of utilizing fewer FPGA resources, while providing accurate downconversion to under a kilohertz. The following sections of the firmware perform the three basic functions of a digital down converter: Down conversion, low-pass filtering and downsampling. 

To downconvert each channel, the $i/q$ time stream is multiplied by the $I_\textup{DDC}$/$Q_\textup{DDC}$ components of the DDC LUT. The DDC LUT is composed of pre-calculated FFT beat frequencies for each channel, sampled at the filterbank-bin sample frequency. Two consecutive channels are operated on in parallel. A single cycle of the operation involves performing the calculation: ($i$ + j$q$)($Q_\textup{DDS}$ + j$I_\textup{DDS}$). Here, $i/q$ are of data type 18.17, $I_\textup{DDC}/Q_\textup{DDC}$ are 16.15, and the resulting $i/q$ output is 19.17. FFT bins containing multiple channels are downconverted once per channel.

For successful down conversion, the DDC LUT playback must be synchronized with the incoming channelizer $i/q$ stream. If no intervention is taken, on system start, the first channel arriving at the down converter will be out of sync channel-wise with its corresponding DDC tone by a number of clock cycles between zero and 512. This `DDC shift' is constant for a given image file, but varies by a small number of clock cycles between different compilations. It is set upon each system start by programming a variable delay block via software input. 

The value of the DDC shift can be determined using a variety of methods. We have found that it is preferable to use snap block data for this purpose, since this same data can also be used to verify that the downconversion is working properly. One successful method of determining the shift, which can be automated in software, is to step through each possible DDC shift using a variable delay block while monitoring the snap block data of a single channel. In software, an FFT is taken of the snap block data at each shift, which includes $i$ and $q$, as well as $I_\textup{DDS}$ and $Q_\textup{DDS}$. When the delay has been set properly, the DDC channel frequency will match that of the $i/q$ time stream, and FFT-bin index for each LUT will be identical.

\subsubsection{Accumulation and Downsampling}

The low-pass filtering and downsampling stages of the DDC are achieved by channel-wise accumulation of $i$ and $q$. The length of the accumulation is set by user input via a software register, which determines the readout bandwidth of the ROACH board. For BLAST-TNG, the accumulation length is set to $2^{19}$ clock cycles, corresponding to 1024 FFTs per accumulation, for a readout frequency of 488.28~Hz.

Two consecutive channel outputs from the down-converter are accumulated independently, as $i$ and $q$, in CASPER vector accumulator blocks of length 512, and permitted to grow to 32-bits. The averaging function of the accumulator is effectively a box-car filter, which provides low pass filtering of the FFT-bin time streams. The division required to complete the averaging of each data sample is performed in software. Following accumulation, the $i/q$ stream is prepared for UDP packetization. BLAST-TNG's scanning strategy requires a nominal readout frequency of 488~Hz, which, for a packet containing 1024 channels, corresponds to a data rate per module of $\sim$4~MB/s. 

\subsubsection{UDP Packetization and Time Stamping}

UDP packetization is performed in firmware using the CASPER one-gigabit Ethernet (1GbE) block. The source MAC, IP address and UDP port for each readout module are hard coded into the block before compilation, whereas the destination IP and port are user configurable. Each UDP frame, including its 42 byte header, are constructed within the 1GbE block. Since the input data width for this block is one byte, each 4 byte sample (either $i$ or $q$) must be sliced byte-wise before being input to the block. After slicing each sample, 8192 bytes of data are input to the block for each frame.

Each data packet is tagged with a coarse and fine time stamp relative to a pulse-per-second (PPS) input fed into the sync-in port on the ROACH-2 board. The PPS is synchronized using the flight computer's GPS system. The coarse count is the number of elapsed PPS pulses since initialization via a user input, and the fine count is relative to a clock cycle counter which is reset by each PPS pulse. The number of elapsed clock cycles since the zeroth count are appended to the packet as the fine time stamp, with provides a time resolution of $\sim$4~ns. Rather than inserting the time stamp values into the UDP header, they are tagged onto the last two channels of the data packet, which are otherwise unused. In the final version of the flight firmware, a 4-byte CRC checksum will also appended to the end of each data packet.  

\section{Software and Flight Operation}\label{sec:software}

A software interface for the ROACH-2 firmware and analog electronics is used to perform critical functions required during flight, which include uploading the firmware to each FPGA, initializing software registers such as the DDC delay, writing and programming carrier frequency combs, acquiring snap block data and programming the attenuators as needed. KATCP facilitates communications between the PPC and either a local DAQ or FC. KATCP can upload firmware to the ROACH-2 flash memory, program it onto the FPGA, read and write to the various software registers, and monitor the temperature sensors on individual ROACH board components. This set of commands forms the basis for controlling the firmware. Higher level functions, such as programming the LUTs or tuning the LO, are performed with either Python or C-based software.

\begin{figure}[h]
\centering
\includegraphics[width=\linewidth]{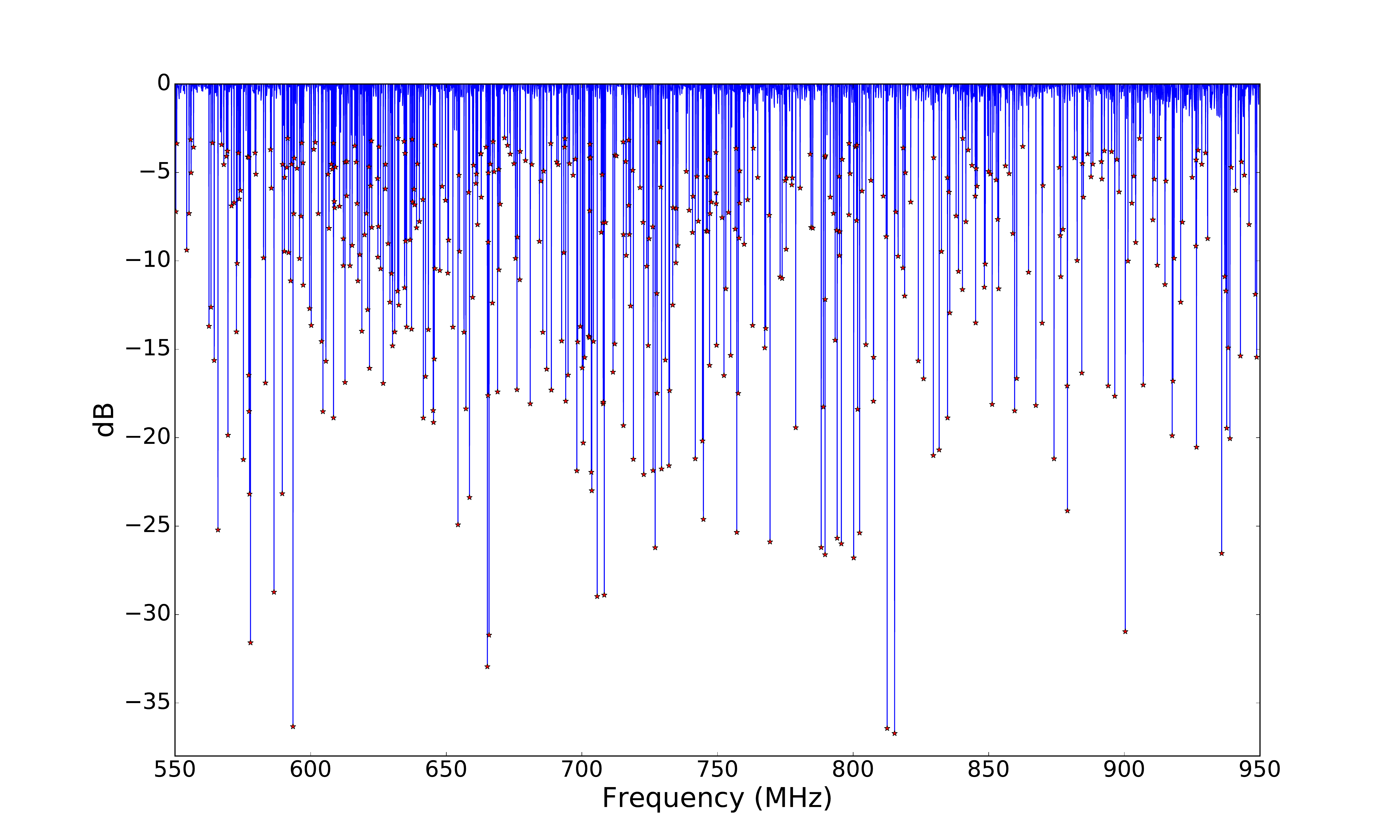}
\caption[An S21 trace of the prototype 250 $\mu$m detector array]{An S21 trace of the prototype 250 $\mu$m detector array, with the locations of resonant frequencies marked by red stars. In this figure, the continuum has been low-pass filtered and normalized to 0 dB, as part of the software algorithm used to identify the resonances.\label{aba:fig5}}
\end{figure}

\begin{figure}[h]
\centering
\includegraphics[width=0.7\linewidth]{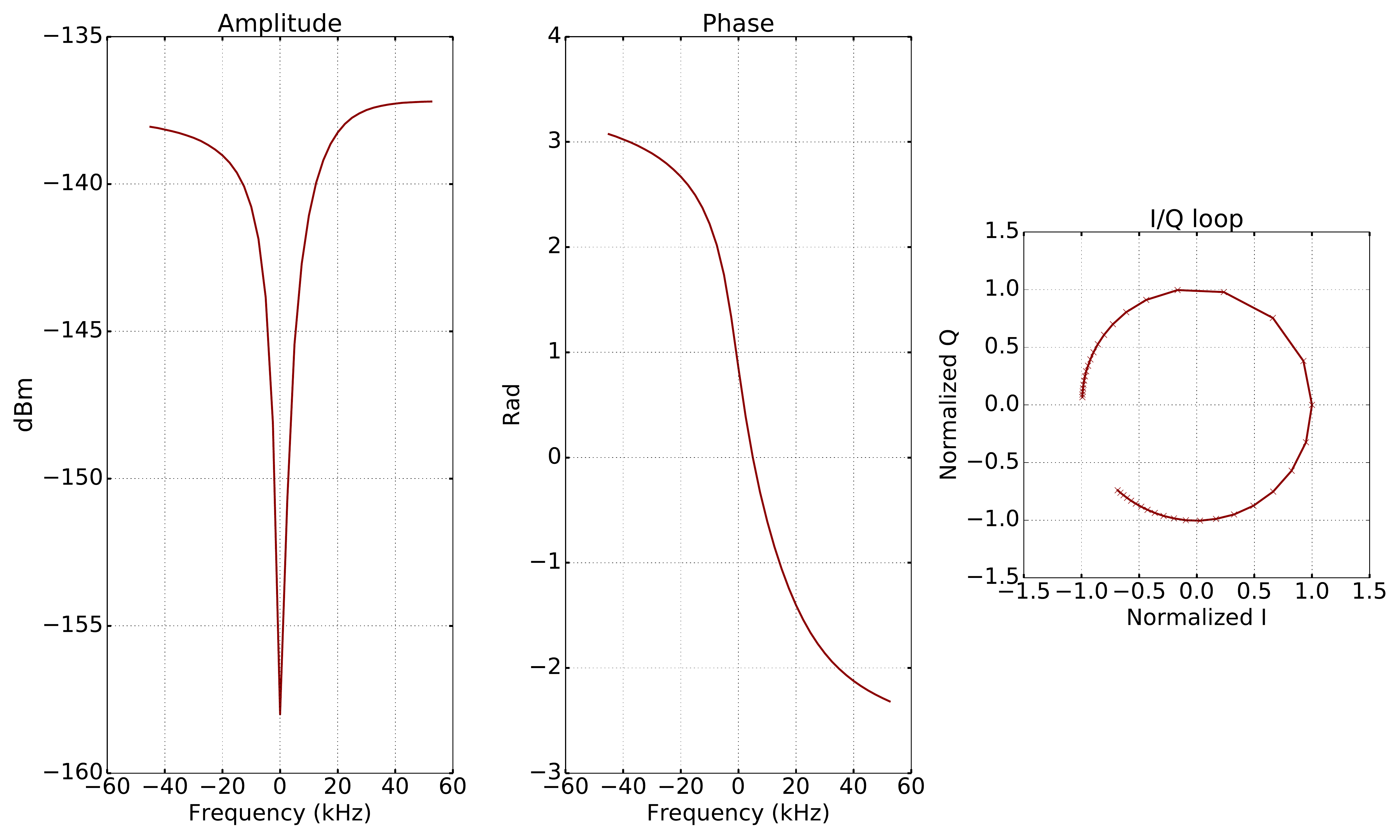}
\caption[Multiplot]{'Target sweep' data reduced to the amplitude, phase, and I/Q loop for a single resonator in the BLAST-TNG prototype 250 $\mu$m KID array. \label{aba:fig6}}
\end{figure}

While the readout section of the FC software is written in C, its essential functions are ported from Python software which has been developed for lab-based ROACH-2 KID readout in conjunction with BLAST-TNG. Software operations begin with the generation of a probe comb containing 500-1000 evenly spaced tones, each having a random phase, which cover the 512~MHz of detector bandwidth. The amplitude of the DAC LUT waveform is normalized to utilize the full dynamic range of the DACs. 

Next, an S21 trace of the array is calculated by stepping the LO through 2.5~kHz increments\footnote{This is the smallest step size permitted by the Valon 5008 Synthesizer.} over the spacing between probe tones in the comb. At each step, a small number of data packets is stored, and the magnitude of $I$ and $Q$ can be plotted as a function of frequency, revealing the location of the resonances in each LeKID array. A peak finding algorithm is applied to the trace to identify the resonant frequency of each detector (Figure \ref{aba:fig5}).

When multiple probe tones fall within a single FFT bin, we choose the two resonances with the highest Q. The list of resonant frequencies is then used to program new DAC and DDC LUTs into each readout module. Next, a narrower 'target sweep' is performed by stepping the LO across 100~kHz of bandwidth centered on each resonance. The target sweep data is saved to disk for characterizing each pixel's phase and amplitude response (Figure \ref{aba:fig6}). This data is also used to calculate each pixel's $I/Q$ gradient ($dI/df$, $dQ/df$) \cite{monfardini14}, a figure of merit used during flight to determine detector loading. After completing the two LO sweeps, the readout is set to data streaming mode, and the telescope may begin science observations. During flight, target sweeps will be regularly repeated to monitor changes in loading on each pixel. 

\section{System noise characterization}

In RF loopback mode, the output of the modulator is attenuated, amplified by the second stage amplifier, and attenuated again before entering the demodulator. The lowest achievable power spectral density (PSD) is set by the performance of the ADC, which, at a sample frequency of 512~MHz, has a phase noise floor of $\sim{-147}$~dBc/Hz \cite{mchugh12}. The crest factor of the DAC waveform also contributes to the PSD. As the number of probe tones in the comb increases, the DAC output power per tone decreases linearly, and the crest factor increases. By assigning a random phase to each tone in a comb containing $N$ carriers, the crest factor is reduced to order $\sqrt{\log_{10}(N)}$ \cite{Boyd86}. In practice, the actual crest factor will be higher. 

\begin{figure}[h]
\centering
\includegraphics[width=0.7\linewidth]{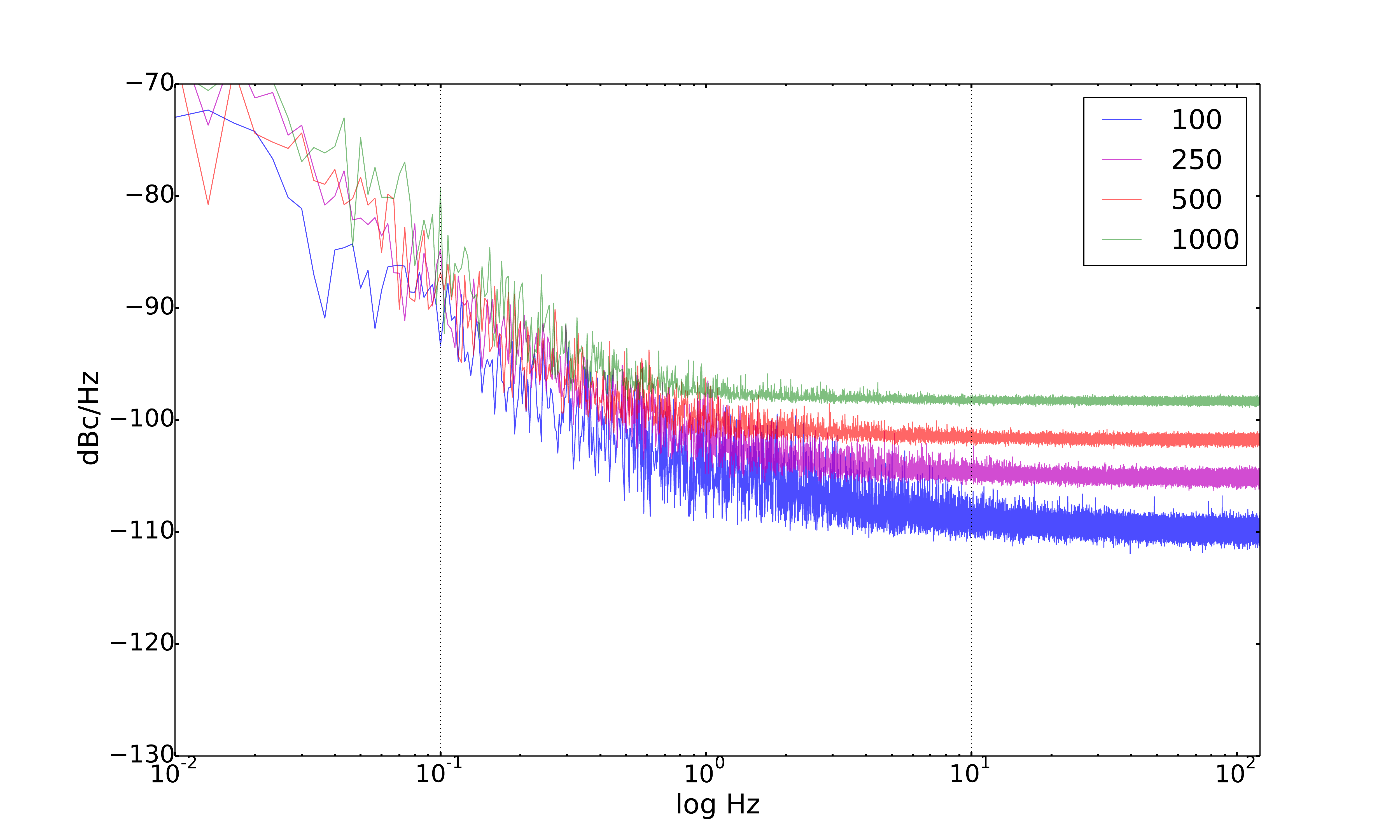}
\caption[]{Phase PSD in RF electronics loopback mode for 100 (blue), 250 (purple), 500 (red) and 1000 (green) channels, acquired over 300 seconds at a readout frequency of 244.14 Hz. The PSD shown is an average over all channels. \label{aba:fig7}} 
\end{figure}

\begin{figure}[h]
\centering
\includegraphics[width=0.7\linewidth]{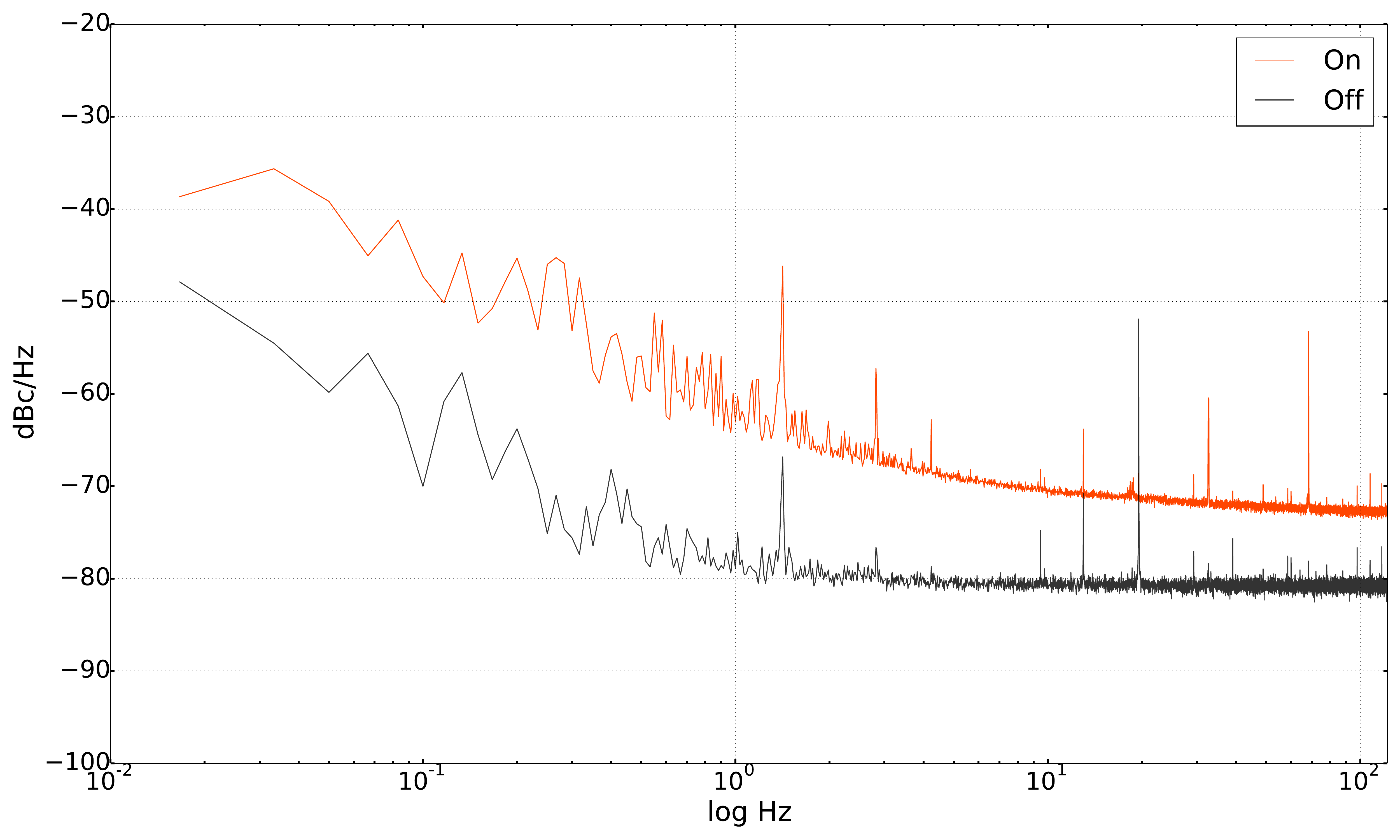}
\caption[]{The average phase PSD measured over 60 seconds for 574 resonators on the BLAST-TNG 250 $\mu$m KID array at NIST, in January, 2016. The PSD is shown on resonance (orange), and 300~kHz off-resonance (black). The spikes in the off-resonance data are likely due to harmonics of AC line pickup and pulse tube cycling at frequencies of a few hertz.\label{aba:fig8}}
\end{figure}

Following an analysis similar to that found in \citet{vanrantwijk16}, the phase PSD which can be expected for a probe comb containing $N$ carriers is calculated as: 
\begin{equation}\label{eq:whitenoise} -147~\textup{dBc/Hz} + 10\log_{10}(N) + 10\log_{10}(\log_{10}N) \end{equation}
The phase PSD as measured in RF loopback mode with probe combs of different $N$ is shown in Figure \ref{aba:fig7} for $N$ of 100, 250, 500 and 1000. In these measurements, data was acquired over a period of 300 seconds at a readout frequency of 244.14 Hz, and the phase is calculated relative to the origin of the I-Q plane. The PSDs are shown as an average over all carriers. For $N =$ 1000, we measure a white noise level of $\sim{-98}$~dBc/Hz, which, by Equation \ref{eq:whitenoise}, indicates a DAC crest factor of $\sim$19~dB. The white noise level is seen to vary linearly with $N$, and the 1/f knee is $\lesssim{0.5}$~Hz. 

A readout module was tested at NIST on the BLAST-TNG prototype 250~$\mu$m detector array in January, 2016. These tests were conducted while the array was dark and held at $T_\textup{bath}$ $\approx$ 50~mK. A 60 second segment of data was collected simultaneously for 574 channels, at a readout frequency of 244.14 Hz. The LO was then shifted by 300~kHz, and another 60 seconds of data were taken. The results are shown in Figure \ref{aba:fig8}, displayed as an average over all channels. The average off-resonance white noise level is less than on-resonance, as is the approximate 1/f knee. In future measurements leading up to the flight, blind tones will included in the probe comb in order to remove the common mode component of the flicker noise.  

\section{Conclusion}

We have developed a new, highly frequency multiplexed readout for large-format superconducting detector arrays based on CASPER's ROACH-2 platform, and designed for use in the upcoming NASA BLAST-TNG balloon-borne mission. The system can readout up to 1024 channels over 512~MHz of instantaneous RF bandwidth centered at 750 MHz, at a readout frequency of 488.28~Hz. At the time of this publication, the system has been shown to provide simultaneous detector-noise limited readout of order $\sim$600 LeKID detectors. The readout hardware and electronics have undergone thermal analysis and testing, and have been deemed flight-ready. 

Analysis of the data from the upcoming December, 2017 flight of BLAST-TNG will provide information about the configuration of the galactic magnetic fields over a wide range of spatial scales not yet accessed by previous experiments, and will help to clarify their role in star formation. In addition, the BLAST-TNG data will provide the first detailed measurements of the variation in the properties of the polarized thermal dust emission across entire Giant Molecular Clouds.

\section*{Acknowledgments}

This work was made possible by the infrastructure created by CASPER and its contributors, and was greatly accelerated through collaborations with researchers at UC Santa Barbara, Caltech/JPL, NIST, Columbia University, University of Pennsylvania, Cardiff University and elsewhere.  

BLAST-TNG is funded by NASA through grant number NNX13AE50G. Detector development is supported in part by NASA through NNH13ZDA001N-APRA. Sam Gordon was funded by a NASA Earth and Space Science Fellowship, NNX16AO91H. Brad Dober was funded by a NASA Earth and Space Science Fellowship, NNX12AL58H. The BLAST-TNG collaboration would like to acknowledge the Xilinx University Program for their generous donation of five Virtex-6 FPGAs for use in our ROACH-2 readout electronics.

\end{document}